\documentstyle[prb,aps,twocolumn,epsfig]{revtex}
\begin{document}
\draft
\tighten
\wideabs{
\title{Estimation of matrix element effects and determination of the Fermi surface in BSCCO systems using angle-scanned photoemission spectroscopy}
\author{S. V. Borisenko \cite{imp}, A. A. Kordyuk \cite{imp}, S. Legner, C. D\"urr, M. Knupfer, M. S. Golden, J. Fink}
\address{Institute for Solid State Research, IFW Dresden, P.O. Box 270016, D-01171 Dresden, Germany}
\author{K. Nenkov and D. Eckert}
\address{Institute for Metallic Materials, IFW Dresden, P.O. Box 270016, D-01171 Dresden, Germany}
\author{G. Yang, S. Abell}
\address{School of Metallurgy and Materials, The University of Birmingham, Birminhgam, B15 2TT, United Kingdom}
\author{H. Berger}
\address{Institut de Physique Appliqu\'ee, Ecole Politechnique F\'ederale de Lausanne, CH-1015 Lausanne, Switzerland}
\author{L. Forr\'o}
\address{DP/IGA, EPFL, 1015 Lausanne, Switzerland}
\author{B. Liang, A. Maliouk, C.T. Lin, B. Keimer}
\address{Max-Planck Institute f\"ur Festk\"orperforschung, Heisenbergstrasse 1, D-70569 Stuttgart, Germany}
\date{\today}
\maketitle
\begin{abstract}
The strong dependence of the momentum distribution of the photoelectrons on experimental conditions raises the question as to whether angle-resolved photoemission spectroscopy (ARPES) is able to provide an accurate reflection of the Fermi surface in Bi-based cuprate superconductors. In this paper we experimentally prove that the main contribution to the intensity variation comes from matrix elements effects and develop an approach to overcome this problem. We introduce a concept of 'self-normalization' which makes the spectra essentially independent of both the matrix elements and particular experimental parameters. On the basis of this concept we suggest a simple and precise method of Fermi surface determination in quasi-2D systems.
\end{abstract}

}

\section{Introduction}
Since the beginning of the field of the high $T_c$ superconductors (HTSC), ARPES has taken a special role in the experimental study of these systems. \cite{OLSON_BOOK,SHEN}
Among the numerous remarkable ARPES experiments on the HTSC, a special place belongs to the investigation of the Fermi surface (FS) of these systems. The difficulties encountered upon applying 'traditional' techniques for determining the FS to the HTSC (such as de Haas - van Alphen and positron annihilation) focussed the attention on the possibility offered by ARPES to obtain a direct image of the basal-plane projection of the FS. 
The vast majority of the earlier ARPES work \cite{SHEN,OLSON_PRB,AEBI,MA_PRB,DING_FS_PRLs,WHITE_PRB} agreed in the view that the FS in the most HTSC is hole-like, and centred at the X,Y points of the
2D Brillouin zone.
Recently, however, a controversy regarding the FS topology of these systems has flared up.
Some groups have suggested the presence of an electron-like, $\Gamma$-centred FS in the Bi$_{2}$Sr$_{2}$CaCu$_{2}$O$_{8+\delta}$ (BSCCO) [Refs. \onlinecite{CHUANG,FENG,GROMKO}] and Pb-BSCCO [Ref. \onlinecite{BOGDANOV}] systems, which would represent a complete revision of our thinking regarding the fundamentals of the electronic structure of the HTSC.
On the other hand, other groups have also re-visited this question \cite{FRETWELL,BORIS,LEGNER,SCHWALLER} and confirmed the 'old' picture of a hole-like FS in the Bi-cuprates. At the same time in La$_{2-x}$Sr$_x$CuO$_4$ a cross-over from hole-like to electron-like FS has been suggested from  ARPES data on going to the overdoped side of the phase diagramme.\cite{INO,YOSHIDA,INO_condmat}
Consequently, there still exists no consensus as to the correct picture for the normal state Fermi surface topology in the HTSC, making this question an important one to clarify.

The current debate as regards the Fermi surface topology is based to a large extent upon ARPES intensity maps.\cite{CHUANG,FENG,BOGDANOV,FRETWELL,BORIS}
This means that the issue of matrix elements, which could be strongly photon energy and {\bf k}-dependent in the 2D cuprate-based materials \cite{BANSIL,HAFFNER,DUERR} has to be treated seriously.
Thus we should be able to identify situations in which the matrix elements dominate and, where possible,
develop practical methods of exctracting the underlying true information from the raw photoemission intensity signal.
A second issue is that of how to accurately determine the Fermi momentum vectors, {\bf k}$_F$ from {\it real}
photoemission data.
The accuracy within which this {\bf k}$_F$ determination can be tested has increased dramatically in the 
past few years as a result of a new generation of electron energy analysers which offer resolutions in \textbf{k}-space 
one order of magnitude superior to what was previously available.

In this paper, we address the question as to how one can best locate Fermi momentum vectors in the HTSC with the aid of the angular distribution of photoemission intensity. As a case study we take BSCCO, but in fact the conclusions arrived at are quite general to the high resolution ARPES investigation of quasi-2D systems. We experimentally demonstrate the strong distortion of the 'pure' photoelectron angular distribution caused by matrix element effects, thus making an appropriate further analysis of the raw data compulsory. To give such a statement a firm foundation, a detailed discussion of the experimental conditions and their influence on the photocurrent together with a critical overview of the existing methods of {\bf k}$_F$ determination are presented. We then demonstrate that, using what we call a 'self-normalization' procedure, one can significantly reduce the dependence of photoemission spectra on the matrix elements and finally show that this approach can be successfully applied to the BSCCO compounds. Consequently on the basis of the self-normalization method we formulate a criterion of determining the Fermi surface of the HTSC from ARPES data.

\section{Methodology}

Because of rapid evolution of the modern ARPES experiment and rising number of possible techniques for the solution of a given problem, we include in this section not only traditional experimental subsection (A) but also a quite detailed description of our approach to the FS mapping. In subsection B we discuss the quantities which are in principle accessible by ARPES in our implementation. Then we discuss how to optimize the experimental parameters for the study of the HTSC cuprates (subsection C) before closing this section with a critical evaluation of the available methods of {\bf k}$_F$ determination (subsection D).

\subsection{Experimental}

Two types of experimental set-up have been used. The majority of the data discussed here were recorded with an overall energy resolution of 19 meV (FWHM) using a SCIENTA SES200 analyser coupled to a high intensity He resonance source (GAMMADATA VUV5000) via a toroidal grating monochromator (giving a degree of linear polarization of ca. 40\%). The SES200 analyser provides an angular resolution of down to 0.2$^\circ$. The single crystals were mounted on a purpose-built, high precision cryo-manipulator which allows the sample to be rotated with a precision of 
better than 0.2$^\circ$ about three perpendicular axes in a wide range of angles. The synchrotron based data were recorded as described in Ref.\onlinecite{LEGNER}.

\subsection{ARPES with analysis of the energy and momentum distributions on an equal footing}

The new generation of electron energy analysers mentioned above have enabled a jump in angular resolution performance as a result of electron optics possessing an angular dispersion capability in the direction parallel to the analyser entrance slit.\cite{MARTENSON} One can visualise the new mode of data collection with the help of Fig. \ref{3D edc+mdc}. This shows the information landing on the analyser's 2D-detector within 7 minutes' measuring time in a {\it static} experiment - i.e. without moving the sample or the analyser. One direction on the detector represents an angular interval and the other direction - an energy interval (in this case $\pm 7^\circ$ and $\sim$0.6eV respectively).

The dataset shown in Fig. \ref{3D edc+mdc} is a 'snapshot' taken along the $\Gamma$-$(\pi,\pi)$ high symmetry direction of the Brillouin zone of Pb-doped BSCCO at 120 K, with the photoemission intensity plotted as a function of both binding energy and momentum. The left panel shows cuts of the intensity distribution $I({\bf k}, \omega)$ parallel to the energy axis, i.e. energy distribution curves (EDC), which are uniquely defined by the fixed value of momentum. One can also cut the same $I({\bf k}, \omega)$ distribution parallel to the momentum axis. These cuts are shown in the right panel and are termed momentum distribution curves or MDCs.\cite{VALLA} An MDC should reflect the vector nature of the momentum, and thus is uniquely defined by the chosen frequency (binding energy) and an arbitrary {\it path} in a two-dimensional {\bf k$_{||}$}-space. It is clear from Fig. \ref{3D edc+mdc} that the modern ARPES machinery enables the simultaneous measurement of the energy and angular distribution of the photoelectrons leaving the sample and therefore permits a treatment of the energy and momentum dependence of the photocurrent on an equal footing.\cite{VALLA,KAMINSKI}

The use of a grey scale to represent the intensity enables an efficient presentation of the three dimensional data set shown in Fig. \ref{3D edc+mdc}. The result, shown in the upper panel of Fig. \ref{edc,mdc,edm,mdm} is called an energy distribution map (EDM). An EDM can be thought of as an array of EDCs (or MDCs) taken along particular path in the BZ within particular range of binding energies. If we now fix binding energy and grey-scale-code the intensity in a series of MDCs covering an area in (k$_x$, k$_y$)-space together, then we arrive at a momentum distribution map or MDM, which represents a constant energy surface. The lower panel of Fig. \ref{edc,mdc,edm,mdm} shows such an MDM for $E$=0 eV binding energy covering a part of the first Brillouin zone of Pb-BSCCO. In our case, the sample rotation involved in the recording of an MDM is such that the individual MDCs representing the angular breadth of the 2D detector are radial in nature, with the origin at the $\Gamma$ point. 

The informative capacity provided by such a map is easy to estimate, although not only the $E_F$-MDM is important. The dataset still possesses the binding energy axis, and so a series of MDMs then represents the evolution of the momentum distribution of the electronic states when going from the Fermi energy towards higher binding energies.\cite{MOVIE}

The inset to Fig. \ref{edc,mdc,edm,mdm} summarizes the completeness of the information available in our ARPES experiment. Here we illustrate the three dimensional (k$_x$, k$_y$, $\omega$)-space which can be probed with high E and {\bf k} resolution. The fourth dimension here is symbolised by the grey scale and represents the photoemission intensity. Recording the ARPES intensity while moving along any vertical direction, i.e. parallel to the energy axis \cite{k-window} will give an EDC, whereas an MDC is the intensity distribution along the arbitrary path which belongs to any of horizontal planes in this space. Horizontal planes themselves are MDMs and the vertical surface (plane) defined by a given path (line) in an MDM is the EDM. Thus, the portion of ({\bf k}, $\omega$)-space shown in the inset to Fig. \ref{edc,mdc,edm,mdm} is confined by three EDMs (one is not visible) and two MDMs (one is not visible) and consists typically of approximately 100000 data points. 

The foregoing discussion has illustrated the potential offered by our experimental setup. Nevertheless, despite significant advances there remain factors such as the resolution or lifetime of the sample surface which make it necessary to optimize the other experimental conditions for the treatment of the physical problem at hand. Moreover, given the widespread use of intensity maps in the literature regarding the FS topology of the HTSC, it is evident that the chosen experimental conditions (such as the experimental geometry, the excitation energy, the photon polarization, the temperature etc.) could decisively change the final picture obtained. Therefore, in the next sub-section we briefly deal with the different experimental parameters which could strongly influence the photoemission intensity distribution in ({\bf k}, $\omega$)-space.

\subsection{Factors influencing the $I({\bf k},\omega)$ distribution of photoelectrons}

In order to discuss different parameters affecting the measured photoemission intensity, we first write an expression for the photocurrent as a function of ({\bf k}, $\omega$)-space.
For quasi-2D systems and under the assumptions that the 'sudden approximation' applies \cite{HUEFNER} and that only a single initial state is involved, the photocurrent can be written in the following form 
\begin{equation}
I({\bf k},\omega) =G_{\bf k}\{M({\bf k}) [A({\bf k},\omega) f(\omega)] \otimes R_{\omega, \bf k} + B(\omega)\},
\label{E1}
\end{equation}
where $G_{\bf k}$ is a mainly geometrical prefactor which will be described below, $M$ represents the square of the matrix element linking the initial and final states, $A$ is the single particle spectral function, $f$ is the Fermi function and $R_{\omega, \bf k}$ is the energy and momentum resolution function.
$B$ is the background, which contains extrinsic effects such as inelastic scattering of the photoelectrons. As an approximation we assume a negligible {\bf k}-dependence of the extrinsic background and likewise a negligible $\omega$-dependence of both the matrix elements and prefactor $G_{\bf k}$ within the energy interval of interest ($\sim$ 0.3 eV).
Eq. 1 makes it clear in a formal manner, that the measured signal is not simply the spectral function, and thus that a number of parameters must be known before $A$ can be extracted.

The pre-factor $G_{\bf k}$ describes the combined effects of extrinsic parameters such as occur upon the rotation of the sample with respect to the analyser (i.e. changing effective photon density in the area of the sample 'seen' by the analyser) or the inequal efficiency of the different channels of the parallel detection system. The raw-data MDM shown in Fig. \ref{edc,mdc,edm,mdm} (lower panel) illustrates the effect of $G_{\bf k}$, as it can be seen that at the interjoins of the two separate arcs of individual (radially arranged) MDCs there is an intensity misfit, mainly due to inequal detector channel efficiencies. In order to minimize the effects connected with $G_{\bf k}$, calibration scans can be carried out by measuring an isotropic photoemitter such as an amorphous gold film. Another, simpler way of overcoming this problem, is the self-normlaization which will be described later.

The generic step-like background $B$ observed in the ARPES of the HTSC is still a puzzle. The authors of Ref. \onlinecite{NORMAN} demonstrated that the contribution from secondary electrons, which could be estimated upon the basis of electron energy-loss spectroscopic data \cite{FINK}, is not sufficient to explain the background which is in correspondence with earlier assumptions in this regard.\cite{LIU} However, our assumption that the background is approximately {\bf k}-independent and of practically negligible intensity at $E_{F}$ in comparison with the main signal is supported by the similar $B(\omega)$ line shape for {\bf k}-points from unoccupied part of the BZ or for those {\bf k} for which spectral function peaks at higher binding energies (e.g. close to the $\Gamma$-point). Therefore, where needed, the background can be safely subtracted. We have found that as a good representative for the $B(\omega)$, an EDC from the vicinity of ($\pi$/2, $\pi$/2) or ($\pi$, $\pi$) points could be taken.

The most important component of Eq. \ref{E1} other than the spectral function itself is the $M$({\bf k}) term describing the matrix elements, which depends upon both the photon energy and the photoelectron momentum via the operator which couples the final and initial state wave functions. The choice of the energy of exciting photons is far from being unimportant even in the case of quasi-2D electronic systems. Firstly, upon changing the photon energy one alters the momentum resolution. Secondly, the 2D-CuO$_2$ plane materials exhibit extremely strong variations in the ARPES intensity of their lowest lying ionisation states as a function of the photon energy.\cite{HAFFNER,DUERR} Thus, in extreme cases, by an unfortunate choice of the photon energy the contribution to the angular distribution of the photoelectrons from a particular initial state can be significantly suppressed.

The same goes for the 'angular' part of the $M$({\bf k}) when using highly polarized radiation. If the experimental geometry can be controlled so as to give a clearly defined symmetry condition - such as can be the case along high symmetry lines in {\bf k}-space - the strong polarization dependence of the photoemission signal can be used as a probe of the symmetry of the initial states involved.\cite{DUERR,DESSAU}
If, however, the strongly polarised radiation is used to measure ARPES spectra away from the high symmetry directions in the Brillouin zone - then the observed photoemission intensity represents only a part of the whole picture.
In this context we note that it has been clearly demonstrated that ARPES intensity maps recorded from BSCCO using the {\it same} photon energy  differ very significantly when recorded with differing polarization geometries.\cite{BIANCONI}

Two ways around this problem spring to mind. Firstly, one could use unpolarized light. The laboratory He source and monochromator used here generates VUV radiation with ca. 40\% linear polarization. Thus, the majority of the MDM intensity is coming from excitation with unpolarised light, meaning that although the polarized component will favour emission from particular states, the global effect is quite small and we are consequently able to 'see' all the states involved. The second possibility is to use the variable polarization offered by modern insertion devices at synchrotron radiation sources to record intensity maps in pairs with complementary polarization geometry.

Moving further through the factors separating a real ARPES experiment from $A$({\bf k}, $\omega$) we come to the energy and momentum resolutions, represented in Eq. \ref{E1} by the function $R_{\omega, \bf k}$. When using an angle-multiplexing analyser, momentum resolution can be projected onto two mutually perpendicular directions - parallel to the entrance slit of the analyser and perpendicular to the slit: $R_{\omega, \bf k}=R_{\omega}R_{k_{\parallel slit}}R_{k_{\perp slit}}$. The resolution parallel to the slit $R_{k_{\parallel slit}}$ is defined by the electron optical characteristics of the spectrometer whereas $R_{k_{\perp slit}}$ is futher controlled by the aperture and entrance slit size. In most cases we used 19 meV $\times0.2^{\circ}\times0.5^{\circ}$ FWHM resolution.

Apart from the angular resolution of the analyser, the flatness of the sample surface as well as the excitation energy and the absolute values of the momenta define the momentum component of the ({\bf k}, $\omega$)-resolution. For example, the {\bf k}-resolution for Fermi level emission in experiments using 'high' photon energies (e.g. 55 eV) is up to a factor 3 worse than while using typical 'low' photon energies (e.g. 21.2 eV as here) for {\bf k}-vectors in the second Brillouin zone of BSCCO for the same angular resolution. 

The practical effect of the energy and momentum resolutions on the measured data depends strongly on the dispersion of the feature in ({\bf k}, $\omega$)-space. The influence of each component increases as the direction of the most rapid change of intensity $I({\bf k},\omega)$ approaches the corresponding axis onto which the total resolution is projected. The direction of the most rapid intensity variation roughly coincides with the normal to the bare band (which is a surface in ({\bf k}, $\omega$)-space). Fig. \ref{Best Edcs} illustrates three exemplary EDCs recorded from pure BSCCO at 40K. In each case at least one of the resolution contributions is zero for our experimental geometry. Fig. \ref{Best Edcs}(a) shows the ${\bf k}_{F}$-EDC taken from the $\Gamma$X cut, for which $R_{k_{\perp slit}}$ is negligible - the width of the feature is defined jointly by $R_{\omega}$ and $R_{k_{\parallel slit}}$. In fact, for the strongly dispersing states along the nodal line ($\approx$ 2 eV{\AA}), the momentum resolution along the slit is the dominating factor. For the data of Fig. \ref{Best Edcs}(a) the energy resolution was 19 meV FWHM and $R_{k_{\parallel slit}}$ was 0.015 {\AA}$^{-1}$ (i.e. $0.2^\circ$). The latter causes an energetic broadening of some 30 meV, which easily outweights the contribution from $R_{\omega}$ itself. In choosing an EDC from the MX cut as shown in Fig. \ref{Best Edcs}(b), we switch off the influence of $R_{k_{\parallel slit}}$. As in the $\Gamma$X case, the momentum resolution (this time $R_{k_{\perp slit}}$) still plays the leading role as regards the instrumental broadening of the observed peak. Fig. \ref{Best Edcs}(c) illustrates an EDC from the M-point, where it is well-known that the sharp peak observed below T$_c$ varies little in binding energy as a function of \textbf{k}, \cite{FEDOROV,CAMPUZANO} meaning that the instrumental contribution to the width of this structure is determined effectively by the energy resolution. The three examples shown in Fig. \ref{Best Edcs} show that the conditions we apply result in a good balance between all three resolution components, when considered in the light of the typical Fermi velocities in the BSCCO-based materials.

Having discussed the influence of all the parameters entering into Eq. \ref{E1}, we conclude that the most unpredictable and therefore difficult factor to deal with which separates the photocurrent from the spectral function are the matrix elements, which, as was mentioned above, are strong in quasi-2D CuO$_2$-plane materials. 

We now turn our attention to a discussion of the best manner in which {\bf k}$_F$ vectors can be derived from the ARPES data of the HTSC. This point is much more than merely a detail of the ARPES data evaluation, and lies at the heart of the current Fermi surface controversy. In particular, given the conclusion that it is only the matrix elements that severely hamper a clear view of the spectral function, our discussion of how to determine {\bf k}$_F$ is thus centered on the question as to which level each method is immune, if at all, from matrix element effects.

\subsection{What is the best way to determine when {\bf k}={\bf k}$_F$?}

\subsubsection{Dispersion method (EDC maximum)}

The relatively coarse {\bf k}-mesh available in the majority of earlier ARPES investigations meant that FS crossings
could only be located by the analysis of a series of EDCs (i.e. an EDM) containing a dispersive feature. The most simple and intuitively direct method is then to follow the energy position of the EDC maximum and to extrapolate obtained dispersion relation to E = $E_F$. This procedure, however, suffers from a number of drawbacks.
Firstly, the influence of the Fermi cut-off distorts the picture within ca. 2kT of $E_F$. Secondly, if the self-energy is frequency dependent, following the EDC maxima will lead to the wrong result.
This point can be simply visualised with the help of the three-dimensional plot, shown in the central panel of Fig. \ref{3D edc+mdc}. As the intensity varies along the bare band, the trace formed by joining the maxima of cuts through this object taken parallel to the binding energy axis (EDCs) can never agree with that obtained by joining the maxima of the cuts taken parallel to the momentum axis (MDCs). If the self-energy only weakly depends on {\bf k}, it is evident that the MDC dispersion is much closer to the 'true' dispersion (i.e. the bare one plus the real part of the self-energy) than that from the EDC maxima. As a consequence, even though it does possess the advantages that it is insensitive to the normalization procedure, to the effects of finite energy resolution and is quite robust with respect to matrix elements effects, the EDC method gives only approximate values of {\bf k}$_F$.

\subsubsection{$\Delta$T method}
It has been proposed from ARPES measurements of TiTe$_2$ and from simulations \cite{KIPP} that $E_F$-MDCs shift as a function of temperature in such a way that the difference of such MDCs turns out to be zero only for {\bf k}={\bf k}$_F$. The practical application of this method to the HTSC is blocked by two points. Firstly, everything has to be measured twice (for T$_1$ and T$_2$) with very high {\bf k}-space location precision, with all other parameters being kept equal. This is often impossible due to the finite lifetime of the cleaved surfaces of the HTSC. Secondly, and more fundamentally, the $\Delta$T method {\it cannot} function if the width of the E$_F$-MDCs concerned is temperature dependent. As this is very clearly the case in the HTSC,\cite{VALLA} the $\Delta$T method is invalid in the context of the HTSC. Without wishing to persue this point further here, we refer the reader to the Appendix for the analytical evidence which forms the basis of these statements.

\subsubsection{Symmetrization}
The symmetrization method is based upon an analysis of the lineshape of the result obtained by mirroring the photoemission EDC's around E=$E_F$ and summing up the two spectra for each {\bf k}-point,\cite{MESOT} a procedure that can be described as $I_S({\bf k},\omega) = I({\bf k},\omega) + I({\bf k},-\omega)$. Within this method {\bf k}$_F$ is defined as the point at which the dip at E$_F$ in the symmetrized EDC's ({\bf k}$<${\bf k}$_F$) turns to maximum. It is evident, however, that upon approaching {\bf k}$_F$ the two peaks originated from the $\omega$ and -$\omega$ spectra will approach each other and become indistinguishable, giving a maximum in the symmetrized EDC over a {\it range} of {\bf k}. One can estimate this {\bf k}-range quantitatively by solving the equation $(d^2 I_S ({\bf k},\omega) / d\omega^2)_{\omega = 0} = 0$. For the model spectral function considered in the Appendix, the solution gives {\bf k}-{\bf k}$_F$ $\sim$ 0.02 {\AA$^{-1}$} at 300K. Consequently, the k point at which the dip in the symmetrized EDC's transforms into a peak is shifted away from the true {\bf k}$_F$ introducing considerably larger error than, for instance, the MDC maximum method gives (see below). Symmetrization does possess the advantage that it eliminates the Fermi cut-off from the {\bf k}$_F$-EDC and answers the question whether a FS crossing occured or not in a given EDM even with the presence of strong matrix element effects.

\subsubsection{Maximal gradient of the integrated intensity}

It is well known that even for interacting Fermi systems, {\bf k}$_F$ is characterised by a jump in the momentum distribution, $n({\bf k})$. For finite temperatures one could still, in principle, detect rapid variations in $n({\bf k})$ and estimate {\bf k}$_F$ from the $\max \left| {\nabla _{\bf k}n({\bf k})} \right|$.
It has been proposed \cite{RANDERIA,CAMPUZANORAPID} that the integrated intensity ($I_{int}$) of an EDC could give a measure of $n({\bf k})$ at one particular {\bf k}-point and thus the analysis of a series of EDC's could present an opportunity to estimate {\bf k}$_F$. This method has been applied to different systems, including BSCCO, and is still intensively 
used.\cite{FENG,BOGDANOV,STRAUB,SCHABEL,ZHOU,RONNING}

There exist, however, the following arguments against the $\left| \nabla I_{int}({\bf k}) \right|$ method. Firstly, $I_{int}({\bf k})$ is not equal to $n({\bf k})$, for the same reasons that the raw photoemission intensity is not equal to the spectral function. Secondly, a single EDC does not represent the photoemission intensity for a single {\bf k}-point, but rather for 
a range of {\bf k}-points.\cite{k-window} For lower photon energies (e.g. 21.2 eV), the finite {\bf k}-interval associated with the finite energy width of an EDC can even be comparable with the momentum resolution. Thirdly, the band structure of the system has to be amenable, in the sense that a single state needs to be isolated and well away in frequency from other features. Even if this is the case, the intensity integration should be carried out over all frequencies - in practice however, energy windows varying between 100-600 meV in width are taken for the integration. Taking a narrower window reduces the similarity with $n({\bf k})$, whereas a wider window results in enlargement of the {\bf k}-interval\cite{k-window} and increases the contribution from deeper lying valence band states. A further difficulty arises from the data analysis in that numerical differentiation introduces additional errors. Such transformation of the raw data also produces a set of additional "false" features on the map \cite{SCHABEL} which have to be identified as such and neglected at a later stage. In any case, the quantitative precision of the {\bf k}$_F$ determination is in direct relation to the width of $\left| \nabla I_{int}({\bf k}) \right|$. This width is much broader than, for example, a typical $E_F$-MDC for the 2 eV{\AA}-like dispersive features in the BSCCO compounds. The factors mentioned above make the $\left| \nabla I_{int}({\bf k}) \right|$ method intrinsically inaccurate. In the Appendix, we use simulations to show further that this method can result in substantial systematic errors determining {\bf k}$_F$, which is in agreement with the results of other authors.\cite{KIPP}

\subsubsection{Maximum intensity method (MDC maximum)}

The maximum intensity method, as first introduced for BSCCO in Ref. \onlinecite{AEBI} is based on measuring the $E_F$-MDM. In this case, the photocurrent is recorded only in a narrow energy window centred on the chemical potential, thus enabling the coverage of large areas in momentum space within a relatively short time. 

The physical basis of this method of {\bf k}$_F$ determination is straightforward. One starts with the reasonable assumption that at finite temperatures the spectral function (independent of the model used to describe it) has a peak at 
$\omega$=0 only for {\bf k}={\bf k}$_F$. It then immediately follows from Eq. \ref{E1} that every $E_F$-MDC corresponding to a path in an $E_F$-MDM which crosses the Fermi surface will show a maximum. This property of $E_F$-MDCs has been recognized and successfully applied for {\bf k}$_F$ determination by a number of groups.\cite{BORIS,LEGNER,VALLA,KAMINSKI} For the MDC peak to lie exactly at {\bf k$_F$}, the influence of the $G_k$-factor, the matrix elements ($M({\bf k})$) and resolution ($R_{\omega, \bf k}$) should not be strong enough to shift the peak position of the spectral function. In this context, the essentially symmetric resolution functions can certainly be regarded as harmless.

\subsubsection{Influence of matrix element effects on {\bf k}$_F$ determination}

It is much more tricky to evaluate how strong the dependence on the matrix elements is for a particular
set of experimental conditions.
Numerical calculations of the photoemission intensity including the matrix elements have been carried out, and predict that the matrix elements should have a dramatic effect on the angular distribution of photoelectrons
that would be detected in an ARPES experiment, given an identical underlying Fermi surface topology. \cite{BANSIL}

In Fig. \ref{32,40,50}(a) we show $E_F$-MDCs and in Fig. \ref{32,40,50}(b) the integrated ARPES intensity $I_{int}$ both recorded for a path in {\bf k}-space along the $\Gamma$-M-Z direction in BSCCO compounds. In each case the four panels show data measured using different photon energies,\cite{LEGNER} whereby the 21 eV panel was recorded using radiation from a He resonance source. 

Comparing the shape of the MDCs and $I_{int}$ traces at different excitation energies it is clear that a {\bf k}$_F$ determination method which involves the photoemission intensity such as the maximal gradient of integrated intensity or the MDC maximum methods may function poorly in such a case.\cite{CROSSING} The vertical dashed lines marking the apparent location of Fermi vectors show different results between the two methods, and, more importantly, different results for the same method, depending on the experimental conditions. This holds even for the maximum-MDC method.

This strong dependence of the apparent location of {\bf k}$_F$ on the experimental conditions used to measure
the ARPES data is, in fact, at the root of the current controversy regarding the Fermi surface topology of the 
HTSC.\cite{CHUANG,FENG,BOGDANOV,FRETWELL,BORIS,LEGNER,GOLDEN} At this stage one could even be led to doubt the value of ARPES as a method of determining the Fermi surface in the HTSC. Obviously, there is an urgent need to find a method which is able to accurately reflect the Fermi surface even in the presence of strong matrix element effects. In the following, final results section, we describe an approach which allows not only an estimation of the distorsions caused by matrix elements but also enables a robust and precise determination of {\bf k}$_F$ vectors.

\section{Fermi surface mapping}

From the discussion of the existing methods of {\bf k}$_F$ determination it follows that, in the absence of strong
matrix element effects, the most precise, simple and physically transparent approach is the MDC maximum 
method.
Therefore one natural way forward is to improve this method by 
minimizing its sensitivity to the matrix elements.
The form of Eq. \ref{E1} suggests the possibility of being able to 'divide out'
both the $G_{\bf k}$ and $M(\bf k)$ prefactors, providing the denominator is proportional to their product.
Within the energy range under consideration, one can take both $M(\bf k)$ and $G_{\bf k}$ to be frequency independent.
Thus, a perfect candidate for such as division would be a signal from the same {\bf k}-point, i.e. 
from the same EDC. At the same time, the reference signal should be a slowly varying function of 
momentum in the vicinity of the expected MDM (MDC) maxima.
In principle, we should restrict ourselves to a rather small energy interval, so as give a narrow {\bf k}$_F$-window 
for each EDC.

Although several possibilities exist for such as matrix-element elimination, in the following we concentrate
on the division by the integrated intensity, as this has already been successfully applied to the determination of 
the FS topology in BSCCO systems.\cite{BORIS} 

As discussed above in the context of the $\left| \nabla I_{int}({\bf k}) \right|$ method, the integrated intensity versus {\bf k} proves to be quite
a slowly varying function in BSCCO in the vicinity of the expected Fermi surface crossings in comparison with the 
{\bf k}-dependence of the Fermi energy intensity (i.e. the $E_F$-MDM). Within the framework of Eq. \ref{E1}, the intensity after division, $I_{norm}$ is given by:
\begin{eqnarray}
I_{norm}&=& \frac{I({\bf k},0)}{\int_{\epsilon}I({\bf 
k},\omega)d\omega}= \\
&=&\frac{G_{\bf k}\{M({\bf k}) [A({\bf k},\omega) f(\omega)] 
\otimes R_{\omega, \bf k}|_{0} + B(0)\}}{ 
G_{\bf k}\{\int_{\epsilon}M({\bf k}) [A({\bf k},\omega) 
f(\omega)] \otimes R_{\omega, \bf k}d\omega + 
\int_{\epsilon}B(\omega)d\omega\}},  \nonumber
\label{E3}
\end{eqnarray}
where $\epsilon$ is the energy window of integration, normally chosen between 600 meV and -100 meV. It is easy to see that prefactor $G_{\bf k}$ cancels out immediately without any additional assumptions, thus automatically solving the problem of the detector efficiency calibration. 

As this method involving the division by the integrated intensity has been criticised as being 'unphysical',\cite{BOGDANOV} we now consider the behaviour of the function described in Eq. 2 above in different parts of the Brillouin zone in detail. We consider three regions in the zone: 'definitely occupied'
($E _{\bf k} > \omega _{max}$, where $E _{\bf k}$ is the quasiparticle dispersion and $\omega _{max}$ is the higher binding energy of the integration window), 'definitely unoccupied' ($E _{\bf k} < 0$) and 'close-to-the-Fermi-surface' ($E _{\bf k} \sim 0$) regions, and discuss the contribution to signal at $E_F$ from the spectral function and extrinsic background. The first region contains {\bf k}-values for which the spectral function peaks more than 0.5 eV from the chemical potential. As regards the 'definitely unoccupied' region, the Fermi cutoff means that the signal at $E_F$ from the spectral function disappears faster going into the unoccupied part of the Brillouin zone, and thus we consider the 'definitely unoccupied' part as being located immediately after the FS crossing. In both of these regions, the $I_{norm}$ signal is determined only by the background lineshape. The weak \textbf{k}-dependence of the background means that $I_{norm}$ should be approximately constant in the 'definitely unoccupied' and 'definitely occupied' regions, although  it is somewhat noisy due to the small values of $B(0)$.
In the immediate vicinity of the expected Fermi surface, we can neglect both terms connected with the background, as $B(0)$ is typically an order of magnitude lower than I({\bf k}$_F$,$E_F$) (as can be estimated from, for example, the data  of Fig. \ref{3D edc+mdc}) and the corresponding ratio of the integrated intensities is $\sim$ 0.15.
In this case, the $I_{norm}$ function can be rewritten as:

\begin{eqnarray}
I_{norm}\sim \left\{ \begin{array}{ll}
\frac{B(0)}{\int_{\epsilon}B(\omega)d\omega}\approx const 
& \mbox{if $E _{\bf k} > \omega _{max}$ or $E _{\bf k} < 0$};\\
\frac{[A({\bf k},\omega) f(\omega)] \otimes R_{\omega, \bf 
k}|_{0}}{\int_{\epsilon}[A({\bf k},\omega) f(\omega)] \otimes 
R_{\omega, \bf k}d\omega} & \mbox{if $E _{\bf k} \sim 0$}.
\end{array} \right.
\label{E4}
\end{eqnarray}

These relations show that, in the vicinity of {\bf k}$_F$, the function is independent of both the matrix elements {\it and} the $G_{\bf k}$-factor. It should be noted, however, that in the case of severe suppression of the intensity related to the spectral function, the background contribution becomes substantial and therefore $I_{norm}$ cannot be considered as matrix element independent - even in a first approximation - but still can be used for identifying such a situation.

Although not immediately obvious, the physical meaning of the $I_{norm}$ function is quite transparent. Consider its behaviour along a single cut through the 2D BZ (e.g. that shown in Fig. \ref{3D edc+mdc}). The numerator is simply an $E_F$-MDC and the denominator is, in this case, the $I_{int}$ along $\Gamma$X. We expect the integrated intensity to show a slow drop in the region of {\bf k}$_F$. At the same {\bf k}-region the true, underlying MDC has a sharp maximum exactly at {\bf k}$_F$.
Thus, the division of the narrow Lorenzian-like MDC function by the slowly falling $I_{int}$ function does not even result in a significant shift of the maximum, and makes itself felt only in the asymmetric shape of the renormalized MDC.
In other words, by normalizing the MDC in such a manner, we do not change the position of its maximum, thus 
preserving its most important property as an indicator of {\bf k}$_F$. 
We note that the asymmetry of the normalised MDC peak can act as an additional guide in 
determining whether the FS has been crossed from the occupied part to 
the unoccupied or in the opposite direction. 

Having now discussed the normalised MDC approach on paper, as it were, we now put it to the test.
As was evident in Fig. \ref{32,40,50}, the determination of whether there is a {\bf k}$_F$ vector along the $\Gamma$-M-Z direction in pure 
Bi2212 represents a quite a challenge.
In Fig. \ref{Normalized synchrotron} we show the result of dividing each of the panels of Fig. \ref{32,40,50}(a)
with the corresponding panel of Fig. \ref{32,40,50}(b) - i.e. Fig. \ref{Normalized synchrotron} contains the renormalised
$E_F$-MDCs obtained by division of the raw MDCs by the $I_{int}$ curves.

The major advance here over Fig. \ref{32,40,50} is that now the normalized MDCs are all very similar, which, of course, is logical as there can be only one spectral function for this system. The nearly perfect coincidence between data recorded with a wide range of excitation energies is remarkable, and leaves no doubt that it was the matrix elements which had led to the differences in the raw MDC's seen in Fig. \ref{32,40,50}.
This example shows that in this way we can get an estimate of the influence of the most unpredictable part of the 
{\bf k}-dependent matrix elements - i.e. that part which is not due to the symmetry selection rules. 

Comparison of the raw [Fig. \ref{32,40,50}(a)] and normalized [Fig. \ref{Normalized synchrotron}] MDCs indicates
that the strongest photon energy dependent variations in the raw intensity, $I(\bf k)$, are in the vicinity of the M-point.
For $h\nu=21.2 eV$, $M(\bf k)$ is asymmetric with respect to M-point, whereby emission from the states in the first
BZ is favoured. The same asymmetry holds for $h\nu=32 eV$, which is in agreement with other experiments.\cite{SAINI}
The most noticeable difference for 32 eV photons is the strong suppression of the spectral weight around the M-point itself - i.e. the raw MDC has a minimum at M. There is less pronounced M-point suppression for $h\nu=40 eV$, and for 
$h\nu=50 eV$ the suppression is no longer observed and the asymmetry has changed to favour emission from the 2nd BZ. This last point is a forerunner of the severe matrix elements effects at $h\nu=55 eV$ where intensity from the second zone is completely dominant.\cite{BOGDANOV} Thus, as regards the spectral weight suppression, the {\bf k}-interval corresponding to that part of the MDC which is 'eaten away' by matrix element effects moves from the second BZ to the first when the excitation energy is varied from 21 to 55 eV, thus resulting in highest intensity suppresion at the M point for $h\nu$ ca. 32 eV, as had been predicted from photoemission calculations.\cite{BANSIL} 

The remaining differences between the self-normalized MDCs presented 
in Fig. \ref{Normalized synchrotron} could be due to a number of factors.
For example, the M-point spectral weight suppression can be drastic enough 
(see, for example, Fig.1(a) in Ref. \onlinecite{LEGNER} for $h\nu=32 eV$)
to mean that one is no longer able to distinguish between the real signal and 
the background.
Three of the curves presented in Fig. \ref{Normalized synchrotron} are from
pristine Bi2212, whose $E_F$-MDM is highly complex in the vicinity of the M 
point.\cite{BORIS}
Thus some of the observed fine structure could be due to the 'FS crossing' of the
two diffraction replicas.\cite{FRETWELL}

Here we wish to stress that using the self-normalization method we are able to get 
an impression of the effects of the matrix elements. 
To determine the matrix element itself in a rigorous way, we would need to 
know the spectral function.
In this sense, the self-normalized spectra shown in 
Fig. \ref{Normalized synchrotron} are not equal to $A(\bf 
k,\omega)$ across the whole range of $\bf k\omega$-space but are very close to 
the spectral function for {\bf k} near {\bf k}$_F$.
It is this property of self-normalized MDC's which makes them so suitable for FS mapping.

As a final comment to Fig. \ref{Normalized synchrotron} we note that the universality of the lineshape
of the normalized MDC's along $\Gamma$MZ in the BSCCO-based HTSC - in which no sharp peaks characteristic
of FS crossings occur either side of the M-point - strongly suggests the absence of main
band FS crossings in this direction in the BZ and thereby lends weight to the hole-like FS side in the 
Fermi surface controversy.

The foregoing discussion was limited to the self-normalization procedure using the integrated intensity
as the normalising quantity.
In Fig. \ref{Maps} we show both raw and normalized MDMs recorded 
at room temperature within 3 hours after the cleavage of a Pb-BSCCO (OD 72K) single crystal.

Firstly, dealing with Fig. \ref{Maps}(b), the self-normalized MDM confirms the behaviour expected for the 
function given in Eq. \ref{E4}. Both the 'definitely occupied' and  'definitely unoccupied' regions have approximately
the same intensity (mid to dark grey [red] tone).
On approaching the Fermi surface from the occupied side, the signal is reduced (giving dark areas) as here the 
contribution from the signal at the $E_F$ is still small but the integrated intensity is already quite large.
There are two futher possibilities: if a band crosses the Fermi level (e.g. along $\Gamma$-X) a 
sharp increase in intensity is observed giving a bright feature on the map.\cite{portion}
Alternatively, in the case of the $\Gamma$-M-Z cut (see Fig. \ref{Normalized synchrotron}), the MDC has a plateau near the M-point, 
reflecting the behaviour of the flat band which does not cross the FS but approaches near enough to it to contribute to the signal at
$E_F$. This leads to a fairly uniform intensity (mid grey [red] tone) around the M-point.
Upon careful analysis of the locations of the MDC maxima in both raw and self-normalised maps, we find that there is no detectable
shift between the two datasets, which confirms again that the $I_{int}$ function varies much more slowly than the MDC does in the vicinity of the
FS.
We now turn to Fig. \ref{Maps}(c), showing the self-normalization result using the intensity at high binding energies (in this case 250 meV
below E$_F$) as the denominator.
The self-nomalised MDM in Fig. \ref{Maps}(c) displays all the characteristics of the map shown in \ref{Maps}(b), indicating firstly that the 
high binding energy signal is also sensitive enough to the matrix element effects to enable their elimination for {\bf k} near to {\bf k}$_F$.
Secondly, this points to the soundness of the assumption that the matrix element effects are insensitive to energy on the range
of 0.5 eV.
Thirdly, the good agreement between the self-normalization based upon the integrated intensity or high binding energy denominators
proves the physically sound basis of the former procedure, in contrast to what is claimed in Ref. \onlinecite{BOGDANOV}.
Finally, we return to the BSCCO FS controversy, and point out that Fig. \ref{Maps} shows without a doubt that the hole-like Fermi surface topology,
which can be the only conclusion upon looking at our FS maps, is not a product of the integrated intensity normalization procedure, but is a robust result.

To summarise this section regarding the use of self-normalization to reduce the strong matrix element effects in ARPES 
of the HTSC we can say:
(i) self-normalization conserves all the advantages of maximum intensity method of {\bf k}$_F$ determination 
and comes close to the ideal of a method which delivers robust, precise results even in the presence of strong matrix element
effects
(ii) the denominator used in the self-normalization should be a signal which 'feels' the matrix element effects, but
which varies relatively slowly in \textbf{k}-space \cite{Above-EF}. We have shown both the integrated intensity and the intensity at higher binding energies to be two candidates which function well.
(iii) the effectiveness of this method means that one can overcome the doubts raised earlier as regards the power of ARPES to determine the Fermi
surface: the self-normalized MDM represents directly the {\it Fermi surface map}.

The self-normalization method also implies a formal criterion for the FS determination: a {\bf k}-point in the 2D BZ belongs to the Fermi surface when and only when all possible normalized MDCs in its vicinity, except may be one (to account for inequal intensity distribution along the FS itself), have a local maximum at that point. To take into account the finite resolution of the experiment one would have need to introduce a tolerance angle within which the MDCs could be considered as belonging to the FS.

As regards the extension of the self-normalization method to quasi-2D systems other than BSCCO, we see no limitations in terms of its validity since the method is based on very general grounds. Even in the case of very low Fermi velocities the self-normalization effectively compensates already observable shift of the $E_F$-MDC due to the finite energy resolution, i.e. gives exact locations of {\bf k}$_F$-vectors.

\section{Summary}
In this paper we have discussed the factors which separate the data of a real ARPES experiment from the spectral function, which is
a highly topical subject in the light of the current controversy regarding the ARPES-derived Fermi surface topology in the HTSC.
Based on high quality ARPES data of the BSCCO and Pb-BSCCO systems recorded under a variety of experimental conditions
(photon energies, degrees of polarization), we have suggested a simple method which enables an estimation of the strength of the matrix 
elements effects in the ARPES of the HTSC, which at the same time allows the precise determination
of the Fermi surface even when the matrix elements are strongly {\bf k}-dependent.
In this approach, a self-normalization effectively immunises the momentum distribution curves - whose maxima deliver precisely the Fermi
wavevectors - against the matrix elements and extrinsic factors separating the photoemission signal from the 
spectral function for {\bf k} near to {\bf k}$_F$.
Consequently, the self-normalized momentum distribution map of the photoemission intensity at $E_F$ gives the most
faithful reproduction of the underlying Fermi surface topology achievable from real ARPES data, and thus provides easy 
access to the quantitative analysis of the Fermi surface in these materials.

\section{Acknowledgments}
This work was funded in part by the BMBF (05 SB8BDA 6), the DFG (Graduiertenkolleg 'Struktur- und Korrelationseffekte in Festk\"orpern' der TU-Dresden) and the SMWK (4-7531.50-040-823-99/6). H.B. is supported by the Swiss National Science Foundation and EPFL. We are grateful to G. Reichardt (BESSY GmbH), R. M\"uller and Ch. Janowitz (Humboldt Universit\"at Berlin) for assistance and to Mike Norman for correcting the error in the calculations.

\section{Appendix}

In this section we communicate in a little more detail certain points as regards the question of how best to determine {\bf k}$_F$.
To enable a more quantitative analysis of the strengths and weaknesses of the various {\bf k}$_F$ methods, we have generated
a simulated dataset, based upon a fit to a {\it real} $\Gamma$X EDM dataset from BSCCO.
Starting from equation (1) given earlier and assuming that the problems of matrix elements, detector calibration and
background have been adequately dealt with, we adopt a model in which the photocurrent can be calculated as

\begin{equation}
I(k,\omega) \propto [A'(k,\omega,R_{k}) f(\omega)] \otimes R_{\omega}.
\label{A1}
\end{equation}

To speed up the calculations, we combine the spectral function with the momentum resolution $R_{k}$ in
\begin{equation}
A'(k,\omega,R_{k}) \propto \frac{\sqrt{\Sigma''^2 + R_{k}^2}}{(\omega - \epsilon_{k})^2 + \Sigma''^2 + R_{k}^2}
\label{A2}
\end{equation}

The absence of a strong asymmetry in the $\Gamma$X MDC's results in a straightforward influence of the momentum resolution,
which is in contrast to the influence of the energy resolution, \cite{KIPP} which is taken into account via convolution with
the resolution function $R_{\omega}(\omega) = (R_{\omega} \sqrt{\pi})^{-1} exp(-\omega^2/R_{\omega}^2)$.
For the imaginary part of the self-energy we use the following approximation: $\Sigma''(\omega,T) = \sqrt{(\alpha \omega)^2 + (\beta T)^2}$ with $\alpha = 1$ and $\beta = 2$ ($\omega$ and $T$ in energy units), which, as can
be seen from Fig. \ref{EDMs}, gives a resonable fit to the experimental data.

Fig. \ref{EDMs} shows typical experimental Pb-BSCCO $\Gamma$X EDM's (for 30 K and 300 K) as contour plots (left panels), 
together with the results of the simulation (right panels).
The quasiparticle dispersion $\epsilon_{k}$ includes the effect of the real part of self-energy $\Sigma'(\omega,T)$, but in the region of interest near
to the Fermi level we consider $\epsilon_{k} = v_F k$, where $v_F$ is simply the renormalized Fermi velocity at $\omega = 0$.
We took $v_F$= 2 {eV\AA} from the experimental data.
Having described the basis of our simulations, in the following we analyse how accurate the {\bf k}$_F$ determined
by the different methods is.

\subsection{Maximum MDC method}

To evaluate the precision of this method quantitatively we simulate the $\Gamma$X $E_F$-MDC according to (\ref{A1}) and (\ref{A2}).
The results of this simulation are presented in Fig. \ref{MDCshift}.

The first observation is that the error in determining {\bf k}$_F$ is very small using this method.
For example, for room temperature the error is less than 0.001{\AA$^{-1}$} for an energy resolution of 19 meV as was used in the experiment.
Even for a resolution of 50 meV, the error is maximally 0.007{\AA$^{-1}$}.
Secondly, the shift of the 'observed' {\bf k}$_F$ from the true value is only weakly temperature dependent, which therefore cannot
be considered as an obstacle to the use of the 'maximum MDC' method.

\subsection{$\Delta T$ method}

The original $\Delta T$ proposal \cite{KIPP} is based upon there being a temperature dependence of the position 
of the MDC maxima.
Thus, the applicability or otherwise of the $\Delta T$ method to the HTSC can also be judged from Fig. \ref{MDCshift}.
The first point is that, as discussed above, the T-dependent shifts of the MDC maxima are very small in BSCCO, in contrast to
the case in TiTe$_2$.\cite{KIPP}
Furthermore, Fig. \ref{MDCshift}(a) also shows clearly that there is no common crossing point on the right flank
of the $E_F$-MDC's, which is a result of the temperature dependence of the width of the MDC's, thus making the $\Delta$T
method inapplicable.
Finally, even if one assumes a temperature-independent width of the $E_F$-MDC, the accuracy of the $\Delta T$ method, $\delta k$,
is related to the uncertainty in the determination of the relative intensities of each of the MDC pairs, $\delta I$: $\delta k \sim \delta I / (dI/dk)_{k=k_F}$.  In our case, to reach an accuracy of $3\times10^{-3}$ \AA$^{-1}$, the $\delta I$ would have to be less than 3\% of $I$, which is beyond most present
experimental capabilities. 

\subsection{Gradient $I_{int}(k)$}

Fig. \ref{NOKshift}(a) shows the results of the simulation as regards gradient $n(k)$, in which 

\begin{equation}
\frac{dI_{int}(k)}{dk} \propto \int_{\omega_{min}}^{\omega_{max}} \left[ \frac{dA'}{dk} f(\omega) \right] \otimes R_{\omega}(\omega) d\omega
\label{A3}
\end{equation}

for $\omega_{min} = -0.1$ eV, $\omega_{max}$ = 0.6 eV is plotted for four different temperatures.
None of the maxima ${dI_{int}(k)}/{dk}$ lie on the $\Delta$k=0 line, indicating a systematic error in the determination of {\bf k}$_F$.
Fig. \ref{NOKshift}(b) shows the temperature dependence of this shift away from the true {\bf k}$_F$, $\Delta k$, plotted for different
$R_{k}$ = (10, 30, 60, 100) meV which are equivalent to (3, 10, 20, 33)$\times10^{-3}$ \AA$^{-1}$ or (0.09, 0.27, 0.54, 0.90)$^\circ$ of angular resolution. 
Since our currently best intrumental angular resolution is 0.2$^\circ$, we discuss the curve for an angular contribution of 30 meV.
Here the error at low temperatures is between 0.002 and 0.003 \AA$^{-1}$, which is as good as the maximum MDC method at these temperatures.
For higher temperatures (e.g. for T$>$T* in the HTSC for which the Fermi surface is not gapped), the error from gradient $I_{int}(k)$ has 
risen to 0.008 \AA$^{-1}$, some 8 times higher than the corresponding value for the maximum-MDC method.

\begin{figure}[bthp]
\caption{Middle panel: photocurrent versus binding energy and momentum along the $\Gamma$-$(\pi,\pi)$ direction at 120 K in Pb-BSCCO (UD85K). Left panel: energy distribution curves (EDC) as parallel intensity profiles corresponding to fixed momentum values. Right panel: momentum distribution curves (MDC) as intensity profiles corresponding to fixed energy values.}
\label{3D edc+mdc}
\end{figure}

\begin{figure}[bthp]
\caption{Upper panel: energy distribution map (EDM) from the $\Gamma$-$(\pi,\pi)$ direction in the Brillouin zone of Pb-doped BSCCO recorded at room temperature. Lower panel: momentum distribution map (MDM) of Pb-doped BSCCO, recorded at room temperature (raw data). White horizontal dashed line represents a {\bf k}$_F$-EDC, vertical ones correspond to  an $E_F$-MDC. In both cases the grey scale represents the photoemission intensity as indicated.
The inset shows the three dimensional (k$_x$, k$_y$, $\omega$)-space which is probed in ARPES of quasi-2D systems.}
\label{edc,mdc,edm,mdm}
\end{figure}

\begin{figure}[bthp]
\caption{ EDCs measured at 40 K from the single cleave of pure BSCCO (UD 89K) at different points in BZ illustrating the contributions from different components of the resolution.}
\label{Best Edcs}
\end{figure}

\begin{figure}[bthp]
\caption{(a) $E_F$-MDCs (circles) and (b) $I_{int}$ (squares) (from 500 meV to -100 meV binding energies) from ARPES data recorded along the $\Gamma$MZ direction in pristine BSCCO for different excitation energies: top right, 32 eV; bottom left, 40 eV; bottom right, 50 eV. The top left panel in each case shows analogous data for Pb-doped BSCCO measured using 21.2 eV photons from a He lamp. The apparent {\bf k}$_F$ locations from (a) the MDC-maximum method and (b) the maximum gradient of the integrated intensity are marked with vertical dashed lines.
}
\label{32,40,50}
\end{figure}

\begin{figure}[bthp]
\caption{Normalized $E_F$-MDCs from ARPES data recorded along the $\Gamma$MZ direction
in pristine BSCCO for different excitation energies: top right, 32 eV; bottom left, 40 eV; bottom right, 50 eV. The top left panel shows analogous data for Pb-doped BSCCO measured using 21.2 eV photons from a He source.}
\label{Normalized synchrotron}
\end{figure}

\begin{figure}[bthp]
\caption{MDM's from Pb-BSCCO (OD 72K) measured with 21.2 eV photons with a low  degree of linear polarization.
(a) raw MDM; (b) and (c) self-normalized MDM's using either the integrated intensity (b) or high binding energy intensity (c) as the denominator of the normalization function.
}
\label{Maps}
\end{figure}

\begin{figure}[bthp]
\caption{Contour plots of typical Pb-BSCCO $\Gamma$X EDMs for two temperatures: 30 K (top) and 300 K (bottom).
The left panels show the experimental EDMs, from which a background has been subtracted and the right panels show 
the results of the simulation.}
\label{EDMs}
\end{figure}

\begin{figure}[tbhp]
\caption{(a) Simulation of the $\Gamma$X $E_F$-MDC for four different temperatures with $R_{\omega}$ = 20 meV and $R_{k}$ = 30 meV ($10^{-2}$ \AA$^{-1}$). (b) Shift of the MDC maximum from the true $k_F$ as a function of temperature for different values of the energy resolution.}
\label{MDCshift}
\end{figure}

\begin{figure}[tbhp]
\caption{Simulations of (a) the gradient $I_{int}(k)$ for the $\Gamma$X direction in BSCCO for four different temperatures. $R_{\omega}$ = 20 meV and $R_{k}$ = 30 meV ($10^{-2}$ \AA$^{-1}$) (see Eq. 6) and (b) the shift of the maximum in gradient $I_{int}(k)$ from the true {\bf k}$_F$ as a funtion of temperature for different momentum resolutions.
For details see text.}
\label{NOKshift}
\end{figure}


\begin{references}
\bibitem{CROSSING} It has been shown, however, that a more detailed analysis can still distinguish between the true FS crossings and "false" ones (see Ref. \onlinecite{LEGNER})
\end{references}
\end{document}